\documentclass[preprint,showpacs,showkeys,
preprintnumbers,amsmath,amssymb]{revtex4}
\usepackage{amsmath}
\usepackage{graphicx}

\newcommand{\<}{\langle}
\renewcommand{\>}{\rangle}
\newcommand{\bol}{\boldsymbol}

\begin{document}


\title{Multiscale Lagrangian Fluid Dynamics Simulation\\for Polymeric Fluid}
\author{Takahiro Murashima$^{1,2}$}
\email{murasima@cheme.kyoto-u.ac.jp}
\author{Takashi Taniguchi$^{1,2}$}
\email{taniguch@cheme.kyoto-u.ac.jp}
\affiliation{
$^1$Department of Chemical Engineering, Kyoto University,
Kyoto 615-8510, Japan\\
$^2$CREST, JST, Kawaguchi, Saitama 332-0012, Japan}

\date{\today}

\begin{abstract}
We have developed a simulation technique of
multiscale Lagrangian fluid dynamics
to tackle hierarchical problems relating to historical dependency of polymeric fluid.
We investigate flow dynamics of dilute polymeric fluid by using the
multiscale simulation approach incorporating 
Lagrangian particle fluid dynamics technique
(the modified smoothed particle hydrodynamics) 
with stochastic coarse-grained polymer simulators 
(the dumbbell model).
We have confirmed that our approach is nicely in agreement 
with the macroscopic results obtained 
by a constitutive equation corresponding to the dumbbell model,
and observed microscopic thermal fluctuation
appears in macroscopic fluid dynamics as dispersion phenomena.
\end{abstract}


\maketitle

\section{Introduction}
Polymer rheology is complicated and exhibits various phenomena,
e.g. vortex growth, die swell, or Weissenberg effect,
caused by a collaboration between microscopic dynamics of polymers
and macroscopic fluid dynamics\cite{Boge93}.
In order to investigate such complex fluids,
we usually take either one of the following two approaches\cite{Bird89}.
One is a fluid dynamics approach regarding whole system as a continuum
from a macroscopic viewpoint, 
and the other is a molecular dynamics approach 
from a microscopic viewpoint.
Although there is a significant difference 
among applicable time and lengths scale in these approaches,
we can improve our comprehension of polymer rheology
from the intercommunication between the different hierarchies
through parameters in constitutive equations.
A lot of constitutive equations have been proposed
to investigate variety of polymer rheology,
but they are not always applicable to any kinds of polymeric liquids and
the generalization of them is a considerably difficult issue.
Then, in order to treat more general polymer rheology,
an available approach is to incorporate 
the fluid dynamics approach directly with the molecular dynamics approach 
without constitutive equations.
CONNFFESSIT (Calculation of Non-Newtonian Flow:
Finite Elements and Stochastic Simulation Technique) 
proposed by Laso and \"{O}ttinger\cite{Laso93} is one of the hierarchical
approaches without assuming constitutive equations,
and is a pioneer work in the multiscale simulation field,
communicating with macroscopic finite elements and 
microscopic ensembles of polymers.
However, it is an extremely heavy technique 
due to remeshing process in the finite element method 
to solve macroscopic fluid dynamics.
After updating the displacements of 
particles holding microscopic ensembles of polymers,
it is needed to remesh the finite elements
in order to preserve one-to-one correspondence 
between macroscopic elements and microscopic particles representing the
center of mass of ensembles.
To avoid the difficulty of remeshing, a semi-Lagrangian approach\cite{Phil99} 
has been produced,
where the convection terms are treated using a finite volume scheme,
although this technique is acceptable 
only when the local stress does not depend on the history of flow.
The heterogeneous multiscale method 
with the fully Eulerian approach at macroscopic level\cite{Ren05,Yasu08} 
is sufficiently practical 
when we consider historically independent fluids, 
e.g. simple fluids, or Newtonian fluids.

In order to treat historical dependency associating with
deformations and orientations of
polymer coils and their entanglements,
we develop a fully Lagrangian multiscale simulation method,
in which fluid points embedding microscopic internal degrees of freedom 
move with the flow field.
Some multiscale simulations are similar in the concept to our new method\cite{Laso97,Hali98}. 
However, these methods are involved and somewhat cumbersome,
since they use the Eulerian and Lagrangian schemes alternatively in
their numerical schemes.
Our new simulation method is technically simple.
By using our method, 
we can directly trace the material points retaining their strain and
strain rate history
and easily handle historically dependent fluids.

Here we employ the modified smoothed particle hydrodynamics 
(MSPH) method\cite{Zhan04} as a macroscopic solver.
The MSPH method highly improves its accuracy especially 
at points near the boundary of the domain
as compared with
the conventional smoothed particle hydrodynamics (SPH) method\cite{Mona05}.
As a microscopic polymer simulation, we use the dumbbell model
which is the simplest model of polymer in a dilute polymer solution 
and a standard
one as used in CONNFFESSIT. 
An ensemble of the dumbbell model corresponds to an analytical model,
e.g. the Maxwell model, and therefore we can verify the validity of 
our simulation results as a benchmark.

We would like to mention Ellero et al.'s work
since the first fully Lagrangian multiscale simulation method has been 
accomplished by them\cite{Elle03}.
In their simulation, the normal SPH method has been employed 
to solve the macroscopic fluid dynamics.
The SPH method fundamentally belongs to a kind of difference methods,
therefore
the first derivative of arbitrary function is defined between two smoothed particles. 
However the MSPH method is a variant of least square fitting methods,
the derivative is defined just on a corresponding smoothed particle in
case of the MSPH method.
The MSPH method is rather straightforward to apply to any differential equations 
than the original SPH method.

The paper is organized as follows.
In the next section we explain the multiscale Lagrangian fluid dynamics method.
We review the MSPH method and the dumbbell model there.
In Section \ref{secNume}, the hybrid simulation is applied to study start-up flows 
in a 2-dimensional channel.
Section \ref{secConc} summarizes conclusions.

\section{Multiscale Lagrangian Fluid Dynamics}
Cauchy's continuum equation of motion serves 
as a starting point to consider general fluids:
\begin{equation}
\rho \frac{d \bol{v} }{d t} 
= \bol{\nabla}\cdot\bol{\sigma}+\bol{f},
\label{eq_Cauchy}
\end{equation}
where $\rho$ is the density of fluid, $\bol{v}$ is the flow velocity,
$t$ is time, $\bol{\sigma}$ is the stress tensor, and
$\bol{f}$ is the body force.
A convective term does not appear in the above equation 
because we treat fluids here in a Lagrangian manner.
The stress tensor  
$\bol{\sigma}$ 
is represented by isotropic pressure $p$ and
the deviatoric stress tensor $\bol{\tau}$:
\begin{equation}
\bol{\sigma}=-p\bol{I}+\bol{\tau},
\end{equation}
where $\bol{I}$ is the $d\times d$ identity matrix and $d$ is the
dimension of the space.
Usually, the pressure $p$ is obtainable 
because of the incompressibility condition, 
but the deviatoric stress tensor $\bol{\tau}$ is not, which
arises from
microscopic internal degrees of freedom.
Since the tensor $\bol{\tau}$
fully depends on microscopic states,
the derivation of the tensor is considerably difficult
especially in complex fluid cases such as polymer fluids.
Even so, many constitutive relations for the tensor $\bol{\tau}$
have been found phenomenologically or
on the basis of microscopic consideration of polymer chains\cite{Doi86}.
They are qualitatively useful in applying to polymer processing,
even if they do not hold microscopically detailed structures.
Because, these microscopic details are compressed 
to macroscopic parameters appearing in the constitutive relations.

Polymeric fluids represent stress-strain and stress-strain rate
historical dependency 
coming from microscopic polymer dynamics.
The historical dependence of stress is essential in
several polymeric phenomena
such as strain hardening or shear thinning.
In order to handle the effect of microscopic polymer dynamics 
we need to consider
microscopic details omitted in the above constitutive equations.
The easiest way to incorporate polymer dynamics is 
to perform a molecular dynamics simulation of polymers,
but such a simulation is impractical to investigate rheological
behaviors
due to the shortness of computationally accessible time.
The other effective simulations have already been developed, and
these simulations have been found to represent not only qualitatively 
but also quantitatively sufficient results\cite{Masu01,Shan01}.
Even so, there is still a gap to apply these simulation methods 
to polymeric flow problems since they are specialized to find
the stress-strain and -strain rate relations in a small simulation box.
If we treat such a simulation system so as to give a constitutive relation
under any applied flow,
the polymeric flow problems on which microscopic polymer dynamics gives
essential influence
become solvable
by incorporating the microscopic polymer simulation into a fluid dynamics simulation.
Such a multiscale simulation method will be available within near
future since computer technology is rapidly developing and 
parallel computing environments become easy to available.
Therefore, we are addressing to advance computational techniques to fully utilize
the high performance computing technology. 
The multiscale simulation technique 
become increasingly important in the future.

In order to develop consistently effective techniques for a future
multiscale simulation,
first of all, we demand a Lagrangian method to solve fluid dynamics 
because of its historical dependence coming from 
the microscopic degrees of freedom.
Secondly, highly parallel computing technique is necessary.
We present such a multiscale parallel computing technique.
We solve start-up flow problems of polymeric fluid 
in a 2-D channel as a benchmark test in the section \ref{secNume}.

\subsection{Macroscopic Fluid Dynamics Simulation\label{subsec_macro}}
We mentioned the importance of historical dependence
of microscopic systems incorporated into the macroscopic system 
in the previous section.
In order to simulate non-Newtonian fluids with a nonlinear 
relation between shear stress and strain rate,
we have to solve the Cauchy's equation of motion in a Lagrangian manner.
If we use a Eulerian method to non-Newtonian fluids,
we need to consider the convection of the stress from adjacent grids,
which means the history of fluids is no longer maintained in the
Eulerian manner
without introducing auxiliary field such as orientation tensor field 
and a variable describing the stretch of polymer coils.
In some trivial cases such as stationary flows
it is easy to obtain the stream line and to know the path through which
the fluids flowed in the past. In such cases, we are able to treat the
non-Newtonian fluids even in the Eulerian framework by taking into
account the convection of the stress\cite{Laso93}.
However, their applications are limited to predictable cases.
In usage of Lagrangian methods we can treat more general cases.

Before we solve any differential equations, 
we have to know spatial derivatives of fields, for example $\bol{\nabla\cdot
\sigma}$ 
appearing in Cauchy's equation of motion.
In the Lagrangian picture, the distribution of fluid particles 
which are used as substitutes for Eulerian grids 
does not necessarily have a regular pattern and changes momentarily.
So it is difficult to define the derivative values in the Lagrangian frame.

Using some kernel functions to interpolate between the unstructured
fluid particles,
the SPH method accomplished to obtain 
average values of the derivatives\cite{Mona05}.
Therefore, when we use the SPH method to solve a differential equation,
we have to be aware of using the average values.
The SPH derives the smoothed derivatives of fields
even when the values are drastically changing among fluid particles.

In order to remove the problem mentioned above,
we implement the MSPH technique
based on the Taylor expansion 
of the original SPH averages\cite{Zhan04}. 
By using the MSPH method, 
we are available to obtain first and higher derivatives
instead of the averaged derivatives which are given 
in the original SPH method.
Once we calculate the derivative values on a fluid particle, 
or a calculation point,
we have only to substitute them 
into the differential equations, 
and then we find their solutions 
by using a time integrating scheme, 
for example the velocity Verlet algorithm or the Runge-Kutta method.
The MSPH method do not need any alternations 
of original differential equations,
whereas the SPH method demand further considerations because 
the derivative of fields obtained by using SPH is
the average value of the differences defined between fluid
particles\cite{Elle03}.
The distinction in their treatments
results from an 
approximation accuracy of these methods:
The approximation accuracy of SPH is
the first order while that of MSPH is the second order.

Here we summarize the MSPH method briefly (see ref.\cite{Zhan04} for
more information).
This method can be regarded as a variant of unstructured grid methods
and least square fitting techniques.
Based on the SPH procedure,
an average of discrete field $f(\bol{x}_i)$ on fluid particles $\bol{x}_i$
are defined by weighted sum:
\begin{equation}
\< f(\bol{x}_i) \> 
\equiv 
\sum_j^{\Omega} f(\bol{x}_j) W(|\bol{x}_i-\bol{x}_j|),
\end{equation}
where $W(x)$ is a Gaussian type function of distance $x$ 
within a finite region $\Omega$. 
Originally, field $f(\bol{x})$ is continuous, 
hereafter we use the following continuous representation:
\begin{equation}
\< f_{\bol{\xi}} \>=\int_{\Omega} d\bol{x}f_{\bol{x}} W_{\bol{\xi}\bol{x}},
\end{equation}
where $f_{\bol{x}}\equiv f(\bol{x})$ and $W_{\bol{\xi}\bol{x}}\equiv W(|\bol{\xi}-\bol{x}|)$.
Then we obtain the following set of equations in accordance 
with the usual SPH procedure:
\begin{subequations}
\label{eq_SPH}
\begin{align}
\< f_{\bol{\xi}}\>_{,\alpha}
\equiv
\< \nabla_{\alpha}f_{\bol{\xi}} \> 
&= \int_{\Omega} d\bol{x} f_{\bol{x}}
 \nabla_{\alpha} W_{\bol{\xi}\bol{x}}
,\\
\<f_{\bol{\xi}}\>_{,\beta\gamma}
\equiv
\< \nabla_{\beta}\nabla_{\gamma} f_{\bol{\xi}} \> 
&= \int_{\Omega} d\bol{x} f_{\bol{x}}
 \nabla_{\beta} \nabla_{\gamma}
W_{\bol{\xi}\bol{x}},
\end{align}
\end{subequations}
where $\nabla_{\alpha} X $ denotes a spatial derivative of $X$ with
respect to the $\alpha$-component of $\bol{\xi}$, i.e. $\xi^{\alpha}$.
We assign coordinate axes to
subscript and superscript Greek characters in the present paper.
Substituting $f_{\bol{x}}$ in the above equations \eqref{eq_SPH}
to the following Taylor series of $f_{\bol{x}}$ 
around a reference position $\bol{\xi}$ 
represented as
\begin{align}
f_{\bol{x}}&=f_{\bol{\xi}}+
(\nabla_{\alpha} f_{\bol{\xi}}) (x^{\alpha}-\xi^{\alpha})\nonumber\\
&\quad+\frac{1}{2} (\nabla_{\beta}\nabla_{\gamma} f_{\bol{\xi}})
(x^{\beta}-\xi^{\beta})(x^{\gamma}-\xi^{\gamma}) + \cdots \nonumber\\
&\equiv
f_{\bol{\xi}}+
f_{\bol{\xi}, \alpha} R^{\alpha}
+\frac{1}{2} f_{\bol{\xi}, \beta\gamma}
R^{\beta}R^{\gamma} + \cdots, \quad(R^{\alpha}\equiv x^{\alpha}-\xi^{\alpha})
\end{align}
the next equations are derived up to the second order in $R$:
\begin{subequations}
\label{eq_preMSPH}
\begin{align}
\<f_{\bol{\xi}}\> 
&=
f_{\bol{\xi}}
\< 1 \>
+
f_{\bol{\xi}, \alpha}
\< R^{\alpha}\>
+
\frac{1}{2}f_{\bol{\xi}, \beta \gamma}
\< R^{\beta}R^{\gamma} \>
,\\
\<f_{\bol{\xi}}\>_{, \delta} 
&=
f_{\bol{\xi}}
\< 1 \>_{, \delta}
+
f_{\bol{\xi}, \alpha}
\<R^{\alpha} \>_{, \delta}
+
\frac{1}{2}f_{\bol{\xi}, \beta \gamma}
\<R^{\beta}R^{\gamma}\>_{, \delta}
,\\
\<f_{\bol{\xi}}\>_{, \epsilon \zeta} 
&=
f_{\bol{\xi}}
\< 1 \>_{, \epsilon \zeta}
+
f_{\bol{\xi}, \alpha}
\< R^{\alpha}\>_{, \epsilon \zeta}
+
\frac{1}{2}f_{\bol{\xi}, \beta \gamma}
\< R^{\beta}R^{\gamma}\>_{, \epsilon \zeta}
.
\end{align}
\end{subequations}
We can rewrite these equations \eqref{eq_preMSPH} to the Matrix form;
\begin{equation}
\begin{pmatrix}
\< f_{\bol{\xi}}\> 
\cr 
\< f_{\bol{\xi}}\>_{,\delta}
\cr 
\< f_{\bol{\xi}}\>_{,\epsilon\zeta}
\end{pmatrix}
\!\!=\!\!
\begin{pmatrix}
\<1 \> && \!\!\< R^{\alpha} \>
&& 
\!\!
\frac{1}{2}
\<R^{\beta}R^{\gamma}\>
\cr
\<1 \>_{,\delta}
&& 
\!\!
\< R^{\alpha} \>_{,\delta}
&&
\!\!
\frac{1}{2}
\<R^{\beta}R^{\gamma}\>_{,\delta}
\cr
\<1\>_{,\epsilon\zeta}
&& 
\!\!
\<R^{\alpha}\>_{,\epsilon\zeta}
&& 
\!\!
\frac{1}{2}
\<R^{\beta}R^{\gamma}\>_{,\epsilon\zeta}
\end{pmatrix}
\!\!
\begin{pmatrix}
f_{\bol{\xi}}\cr f_{\bol{\xi}, \alpha} \cr f_{\bol{\xi}, \beta\gamma}
\end{pmatrix}\!\!.
\label{eq_MSPH}
\end{equation}
The number of rows in Eq. \eqref{eq_MSPH}
is seven in two dimensional space, whereas thirteen 
in three dimensional space.
Considering the commutativity of spatial derivatives 
($\nabla_{\alpha}\nabla_{\beta}f=\nabla_{\beta}\nabla_{\alpha}f$),
we can reduce 
the number of rows 
in Eq. \eqref{eq_MSPH} from seven to six in two dimensions 
because of $f_{\bol{\xi},yx}=f_{\bol{\xi},xy}$,
and from thirteen to ten in three dimensions;
\begin{equation}
\begin{pmatrix}
\< f_{\bol{\xi}}\> 
\cr 
\< f_{\bol{\xi}}\>_{,\delta}
\cr 
\< f_{\bol{\xi}}\>_{,\epsilon\zeta}
\end{pmatrix}
\!\!=\!\!
\begin{pmatrix}
\<1 \> && \!\!\< R^{\alpha} \>
&& 
\!\!
C(\beta,\gamma)
\<R^{\beta}R^{\gamma}\>
\cr
\<1 \>_{,\delta}
&& 
\!\!
\< R^{\alpha} \>_{,\delta}
&&
\!\!
C(\beta,\gamma)
\<R^{\beta}R^{\gamma}\>_{,\delta}
\cr
\<1\>_{,\epsilon\zeta}
&& 
\!\!
\<R^{\alpha}\>_{,\epsilon\zeta}
&& 
\!\!
C(\beta,\gamma)
\<R^{\beta}R^{\gamma}\>_{,\epsilon\zeta}
\end{pmatrix}
\!\!
\begin{pmatrix}
f_{\bol{\xi}}\cr f_{\bol{\xi}, \alpha} \cr f_{\bol{\xi}, \beta\gamma}
\end{pmatrix}\!\!.
\label{eq_MSPH2}
\end{equation}
Here we introduce coefficients
$C(\beta,\gamma)\equiv 1-\frac{1}{2}\delta_{\beta\gamma}$
where $\delta_{\beta\gamma}$ is Kronecker's delta.

In order to obtain the differential values 
$f_{\bol{\xi},\alpha}\equiv\nabla_{\alpha}f_{\bol{\xi}}$,
our procedure is as follows.
At first we calculate the average values $\< \cdot \>$ in Eq. \eqref{eq_MSPH2}
in the usual SPH manner,
and then we solve the above linear equation \eqref{eq_MSPH2} 
by using, for example, the LU decomposition.

Here, we adopt a Gaussian kernel function $W(x)$ 
firstly used in the MSPH method\cite{Zhan04}:
\begin{align}
W(x)=\begin{cases}
\frac{A_d}{(h\sqrt{\pi})^d}
\left(e^{-\frac{x^2}{h^2}}-e^{-4}\right), & \mbox{if}\quad |x| \leq 2h\\
0, & \mbox{otherwise}
\end{cases}
\end{align}
where $d$ is a dimensionality of the space and the cut-off radius is $2h$. 
The normalization parameter $A_d$ equals to 1.04823, 1.10081 and 1.18516
for $d=$ 1, 2 and 3, respectively.
The density of each distributed particle $\rho(\bol{\xi})$ is obtained by integrating
the kernel function in the same manner as the ordinal SPH method:
\begin{align}
\rho(\bol{\xi})=\int_{\Omega} d\bol{x} \, m W(|\bol{x}-\bol{\xi}|),
\end{align}
where $m$ is mass of particle.

As a benchmark of our multiscale simulation,
we investigate two types of macroscopic simulation 
with constitutive relations,
Newtonian viscosity model (Newton model) 
and Maxwell constitutive equation (Maxwell model).
Newton model
is symbolically represented by a dashpod with a viscosity constant $\eta$.
The deviatoric stress tensor of Newton model is proportional to strain rate:
\begin{align}
\bol{\tau}=2 \eta \bol{D} 
= \eta \left(
\bol{\kappa} + \bol{\kappa}^{\rm T}
\right),
\label{eq_Newton}
\end{align}
where $\bol{\kappa}$ is velocity gradient
tensor defined as $\bol{\kappa}\equiv\bol{\nabla v}$.
Maxwell model is symbolically represented by connecting
a dashpod with a viscosity constant $\eta$ and a spring with an elastic
constant $G$ in series.
Maxwell constitutive equation is expressed as follows:
\begin{align}
\bol{\tau}+\lambda\dot{\bol{\tau}}=2 \eta \bol{D},
\label{eq_Maxwell}
\end{align}
where $\lambda$ is a relaxation time.
The constants $\eta$ in Eq. \eqref{eq_Maxwell} and $G$ are connected through $\eta=G\lambda$.

Assuming the deviatoric stress tensor $\bol{\tau}$ in the Cauchy's
equation of motion \eqref{eq_Cauchy}
as to be Newton model,
we obtain the Navier-Stokes equation:
\begin{equation}
\rho \frac{d \bol{v} }{d t} 
= -\bol{\nabla}p + \eta \bol{\nabla}^2 \bol{v}+\bol{f}.
\end{equation}
Even though we can solve the Navier-Stokes equation directly by using
the MSPH method, 
we elaborately solve the Cauchy's equation of motion \eqref{eq_Cauchy}
after obtaining the deviatoric stress tensor $\bol{\tau}$
to be applicable to general problems.

We shortly mention how to determine the isotropic pressure $p$.
When we consider high pressure case, e.g. a flow in a duct or shock wave,
we can treat an incompressible fluid by solving the Poisson equation of
pressure $p$ as shown in Ref. \cite{Kosh96}.
In low pressure case, the effect of compressibility is negligibly small
even if we assume a barotropic fluid $p=p(\rho)$.
In this work, we investigate a fluid in low pressure environment only.

\begin{figure}[t]
\begin{center}
\includegraphics[width=12cm]{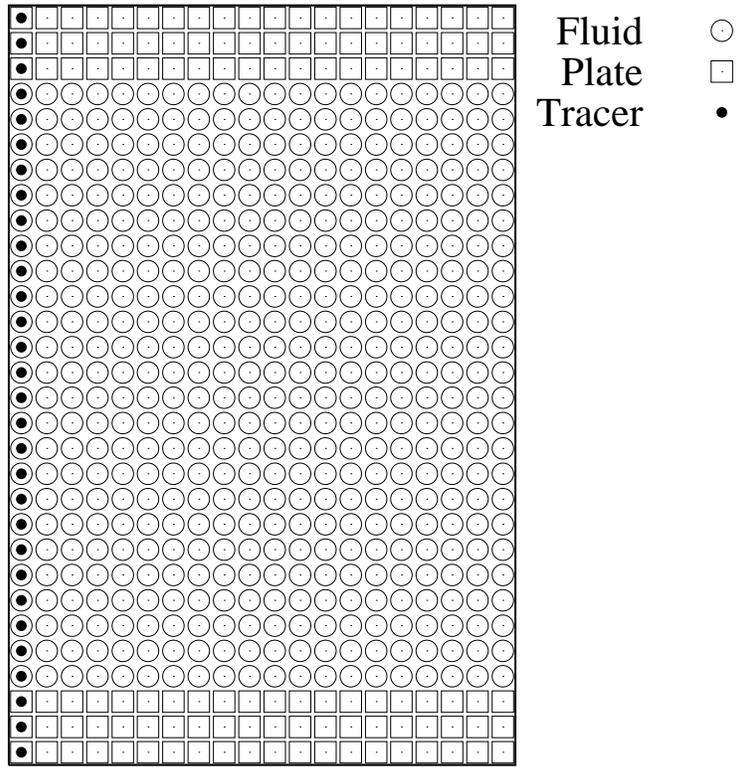}
\caption{Initial configuration for 
Lagrangian fluid simulation.
\label{fig_init}}
\end{center}

\end{figure}

We investigate start up flow profiles obtained by solving the Cauchy's
equation of motion \eqref{eq_Cauchy} with two different types of
constitutive equations, Newton model \eqref{eq_Newton} and Maxwell model
\eqref{eq_Maxwell}.
For convenience, we concentrate on the flow behavior between two parallel
plates.
Assuming translational symmetry for the $z$-direction which is parallel to plates and
perpendicular to the flow direction,
we can treat the present system in two dimensional space.
At first, we prepare a rectangular system box $(Lx\cdot Ly)$
in which particles are regularly displaced
as shown in Figure \ref{fig_init}.
We set
the initial distance $a$ between nearest neighbor particles,
the mass of particles $m$ and
simulation time interval $t_0$ as unity.
The total number of particles $N$ are $L_x\times L_y$.
We set the thickness of plates $l_{\rm w}$ to $3a$ larger than $h$ 
in order to maintain the density of fluid near the plates as much as that of bulk.
We assume the fixed boundary condition for 
fluid particles on plates
and
the periodic boundary condition for the $x$-direction.
Applying the body force $\bol{f}=(1.0\times 10^{-3},0.0)\times m/(a^2 t_0^2)$ to the fluid,
the fluid flows toward the $x$-direction.
Values of simulation parameters are summarized in Table \ref{tab_param}.
The displacements of fluid particles are updated according to the
following equation after solving the Cauchy's equation of
motion \eqref{eq_Cauchy}:
\begin{align}
\frac{d \bol{r}}{d t} = \bol{v},
\end{align}
where $\bol{r}$ is the position of fluid particle.
In order to treat the time-derivative numerically, 
we discretize it as
\begin{align}
\frac{d f}{d t} \simeq \frac{f^{\rm New}-f^{\rm Old}}{\Delta t}
\end{align}
with the finite difference $\Delta t=0.01 t_0$.
We use the velocity-Verlet algorithm as time-integral scheme.

\begin{figure}[t]
\begin{center}
\includegraphics[width=12cm]{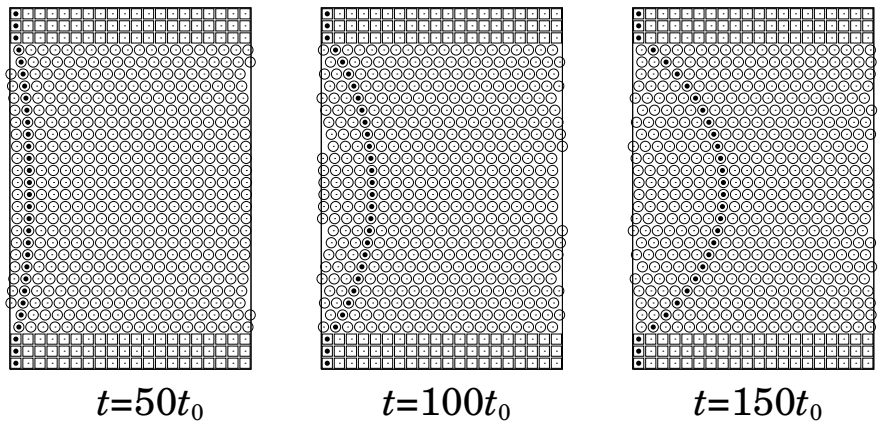}
\end{center}
\caption{Profiles of Newtonian flow
($t=50t_0, 100t_0, 150t_0$).\label{fig_Newton}}
\end{figure}

Figures \ref{fig_Newton} and \ref{fig_Maxwell}
represent the flow profiles in cases of Newton model and Maxwell
model, respectively.
From the positions of tracer particles we can extract
their characters of flow behaviors.
In case of Newton model,
flow profiles show a plane Poiseuille flow,
however, those of Maxwell model represent a plug flow like behavior 
when the elastic constant is very small and the relaxation time is very long.
Maxwell model \eqref{eq_Maxwell} corresponds to Newton model
\eqref{eq_Newton} in the limit $\lambda \dot{\bol{\tau}} \to 0$,
while keeping $\eta$ constant,
which means that
the elastic property of the spring can be negligible.
Therefore, when the relaxation time $\lambda$ becomes smaller,
the flow behavior of Maxwell model approaches to that of Newton
model.

\subsection{Microscopic Polymer Simulation\label{subsec_micro}}
In this study, we select the dumbbell model as microscopic model of
polymer in our multiscale simulation since
the dumbbell model is the simplest and in the statistical limit it
corresponds to an analytical model, i.e. the Maxwell model.
The dumbbell model is modeling a dilute polymeric fluid without
entanglements between polymers.
The dynamics of a dumbbell consists of 
elastic dynamics and thermal fluctuation. 
\begin{align}
\Delta\bol{Q}&\equiv\bol{Q}(t+\Delta t)-\bol{Q}(t)\nonumber\\
&=
\left(
\bol{\kappa}\cdot\bol{Q}
-\frac{1}{2\lambda}\bol{Q}
\right)\Delta t
+\sqrt{\frac{\Delta t}{\lambda}}\bol{\Phi},
\label{eq_dumbbell}
\end{align}
where $\bol{Q}$ is end-to-end vector of the dumbbell,
and
$\bol{\Phi}$ is the Gaussian white noise that is satisfying the
following equations:
\begin{subequations}
\begin{align}
\< \bol{\Phi}(t) \>&=0,\\
\< \Phi_{\alpha}(t)\Phi_{\beta}(t')\> &=\delta_{\alpha \beta}\delta(t-t').
\end{align}
\end{subequations}
We can derive the expression for the dumbbell model \eqref{eq_dumbbell} from microscopic view
points\cite{Otti96,Lars99}.
For simplicity, we only explain about the roles of each term here.
The parentheses of the first term in Eq. \eqref{eq_dumbbell} consists of Affine deformation 
and elastic dynamics.
The relaxation time $\lambda$ is related to the elastic constant
$G$ as discussed in the previous subsection.
When the relaxation time of dumbbell is large, 
the dumbbell is easy to be extended since the elastic constant
$G=\eta/\lambda$ is small. 
The relaxation time of dumbbell also appears 
in the coefficient of second term in the right-hand-side of
Eq. \eqref{eq_dumbbell}
through the fluctuation-dissipation theorem.
The dumbbell with a long relaxation time 
is sensitive to the Affine deformation rather than 
the elastic relaxation and the thermal fluctuation.
The averaged stress in a microscopic simulation system 
is represented by Kramers-Kirkwood formula:
\begin{align}
\bol{\tau}&=G(\<\bol{Q}\bol{Q} \>-\bol{I}).
\label{eq_KK}
\end{align}
Generally speaking, the Kramers-Kirkwood formula consists of
contributions from a solvent,
intramolecular forces, external forces, dynamics of beads, 
and isotropic pressure\cite{Bird89}.
To simplify our arguments, we only consider the effects
of intramolecular forces and isotropic pressure in Eq. \eqref{eq_KK}.
This lead us to concentrate on the orientation and length of 
end-to-end vector of dumbbell.

To update the end-to-end vector according to
Eq. \eqref{eq_dumbbell},
we adopt the second order stochastic Runge-Kutta algorithm here\cite{Bran98}.
This algorithm decreases order of numerical errors from $O((\Delta t)^2)$
contained in ordinary Brownian dynamics algorithms to $O((\Delta t)^3)$.
These errors are caused by its numerical algorithm 
not by the random noise.
Initial conditions of dumbbells are prepared by performing the dumbbell
simulation under a stationary condition over twice of the relaxation time.

\begin{figure}[t]
\begin{center}
\includegraphics[width=12cm]{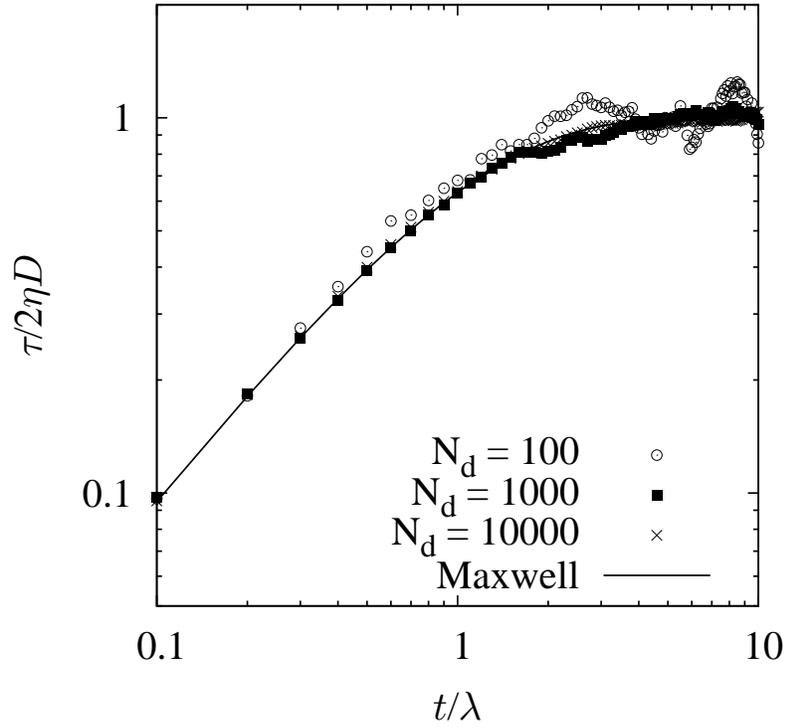}
\caption{The stress histories under the constant shear rate
 $\kappa=1.0/t_0$
for the systems consists of the dumbbells 
with $N_{\rm d}=100$, $N_{\rm d}=1000$, and $N_{\rm
 d}=10000$,
and for the Maxwell constitutive equation.
The thermal fluctuation depends on the number of dumbbells.\label{fig_stress}}
\end{center}
\end{figure}

Figure \ref{fig_stress} shows the stress history 
under a constant shear rate $\kappa=1.0 /t_0$ with $\lambda=100 t_0$ and $G=0.01m/a$
comparing the dumbbell model and Maxwell model.
The thermal fluctuation appears after the Newtonian viscosity $\eta=\tau/\kappa$
grows up to the zero shear viscosity, however it doesn't appear 
or negligibly small in the transient region.
This stress history process can be separated into two processes,
elongation and rotation of dumbbells, respectively.
The border line exists at more or less $t/\lambda=1$.
The stress tensor \eqref{eq_KK} 
is insensitive to the effect of random noise
in the elongation process,
since the orientations of dumbbells are almost same direction.
In the rotation process, however, the angular velocity of dumbbells
is sensitive to the length of end-to-end vector, and therefore
the thermal fluctuation becomes influential.
Increasing the number of dumbbells $N_{\rm d}$, 
the fluctuation of stress is decreasing in proportional to 
$N_{\rm d}^{-1/2}$
by the law of large numbers in statistics
because the stress is defined by 
a statistical average of 
$G(\bol{Q}\bol{Q}-\bol{I})$
over $N_{\rm d}$ dumbbells as shown in Eq. \eqref{eq_KK}.
We expect
the dumbbell model corresponds to Maxwell model 
in the limit $N_{\rm d}\to\infty$.
As shown in Fig. \ref{fig_stress},
the thermal fluctuation has been almost suppressed 
when $N_{\rm d}=10000$.

\subsection{Macro-micro Hybridization\label{subsec_MM}}
In the previous subsections \ref{subsec_macro} and \ref{subsec_micro},
we have introduced the simulation techniques 
of macroscopic fluid dynamics and microscopic polymer dynamics
and explained about their features.
Now we incorporate them in order to treat polymeric fluid dynamics.
The main idea of our hybrid simulation is similar to that of
CONNFFESSIT;
we perform the polymer simulation in stead of the constitutive equation
in order to obtain the stress tensor $\bol{\sigma}$ in Cauchy's equation of
motion \eqref{eq_Cauchy}.

The procedure of our simulation is summarized to the following
steps:
\begin{enumerate}
\item Macroscopic fluid dynamics simulation to update $\bol{r}$,
      $\bol{v}$, and $p$.
\item MSPH method to obtain $\bol{\nabla v}$.
\item Microscopic polymer simulation to update $\bol{\tau}$.
\item MSPH method to obtain $\bol{\nabla \cdot\sigma}$.
\item Return to 1.
\end{enumerate}
We perform these processes every time steps.
We adopt the linked list algorithm\cite{Rapa04} to find a pair of
particles at macroscopic fluid simulation
in order to decrease the calculation time $O(N^2)$ to $O(N)$ for $N$
particles.
In each step, 
we can perform the calculations on a particle independently
from those of other particles,
and therefore we benefit from usage of parallel computing.

\section{Numerical Results on Hybrid Simulation\label{secNume}}
Now we investigate the flow profiles of multiscale Lagrangian fluid
dynamics simulation and compare the resultant flow profiles with the macroscopic results 
shown in Sec. \ref{subsec_macro}.
We perform the multiscale simulation under the same condition 
explained in Sec. \ref{subsec_macro},
however we use the dumbbell model to obtain the deviatoric stress tensor $\bol{\tau}$.
We set the number of dumbbells $N_{\rm d}$ 
in each fluid particle to 1000.
Under the environment using $2\times$3.16 GHz 
Quad-Core Intel Xeon processor (8 cores) 
and OpenMP (Open Multi-Processing) programing interface,
the calculation time for 100 cycles of the simulation procedure 
explained in the previous section
\ref{subsec_MM} is about 2.5 minutes, which corresponds to $t=1.0t_0$ because
of $\Delta t=0.01t_0$.

\begin{figure}[t]
\begin{center}
\includegraphics[width=12cm]{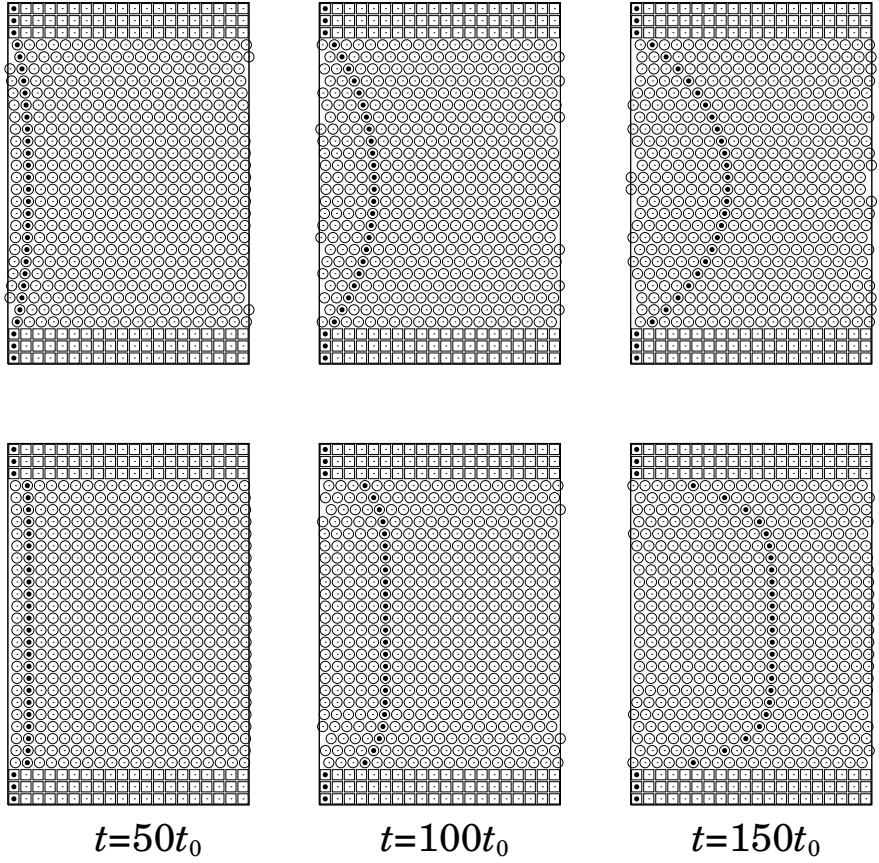}
\end{center}
\caption{Flow profiles in case of Maxwell model ($t=50t_0, 100t_0, 150t_0$) with
$\lambda=10.0t_0$ and $G=0.1m/a$ (upper group), or $\lambda=1000t_0$ and
 $G=0.001m/a$ (lower group). In both cases, $\eta=G\lambda=1.0m/(a t_0)$.
\label{fig_Maxwell}}
\end{figure}

\begin{figure}[t]
\begin{center}
\includegraphics[width=12cm]{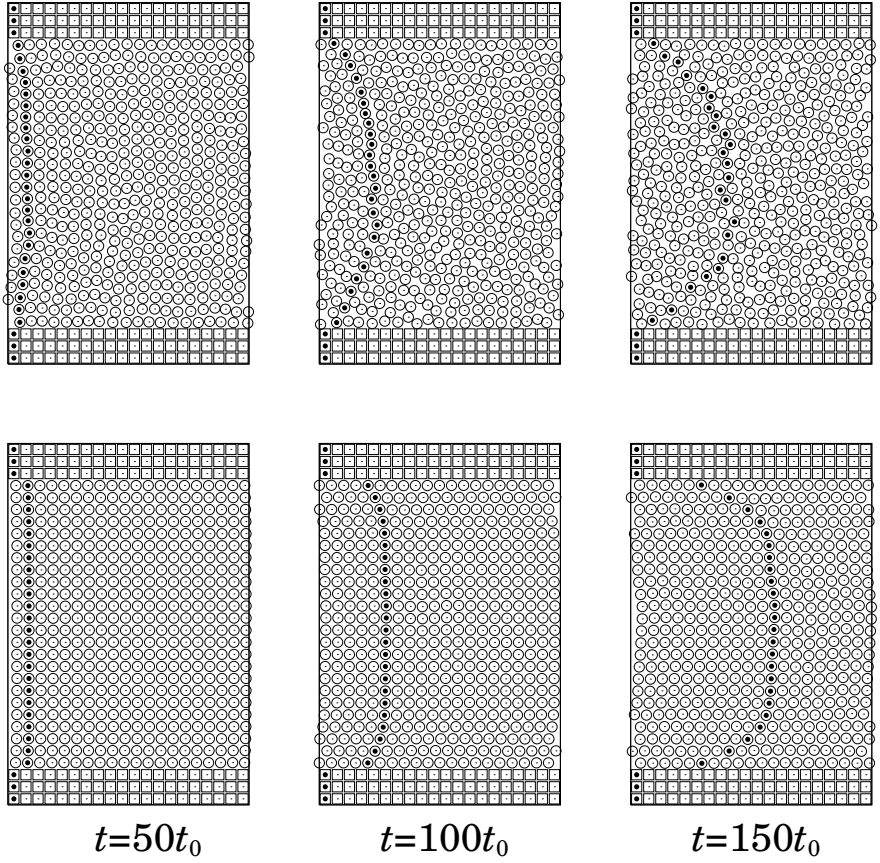}
\end{center}
\caption{Flow profiles obtained by performing
multiscale Lagrangian fluid dynamics 
($t=50t_0, 100t_0, 150t_0$) with
$\lambda=10.0t_0$ and $G=0.1m/a$ (upper group), 
or $\lambda=1000t_0$ and $G=0.001m/a$ (lower group).
Each fluid particle has 1000 dumbbells instead of Maxwell model.
\label{fig_MLFD}}
\end{figure}

Figure \ref{fig_MLFD} represents the flow behaviors obtained from our
new simulation technique.
Comparing between Figs. \ref{fig_Maxwell} and \ref{fig_MLFD},
the tendencies of these flows are almost similar, 
except for the thermal fluctuation observed in Fig. \ref{fig_MLFD}.
When the relaxation time $\lambda$ is small,
the fluctuation of fluid becomes larger.
There are two reasons.
One reason is that the coefficient of random noise in
Eq. \eqref{eq_dumbbell} is large in $\lambda$ small case, 
because of the fluctuation dissipation theorem.
Another one is that
time of elongation process, discussed in
Sec. \ref{subsec_micro}, becomes short,
and then the thermal fluctuation appears earlier. 

We have also performed the multiscale simulation with $N_{\rm d}=10000$.
The resultant flow behaviors have been very close to those of the Maxwell model which
are shown in Fig. \ref{fig_Maxwell}, 
since the thermal fluctuation in this case has been almost suppressed
as mentioned in Sec. \ref{subsec_micro}.

Strictly speaking, 
the incompressibility doesn't satisfied at
some local regions in the present simulation.
In order to improve this failure, we should solve the Poisson equation
of isotropic pressure $p$\cite{Kosh96}.

\section{Conclusion\label{secConc}}
We have developed the multiscale Lagrangian fluid dynamics technique 
to simulate a hierarchical polymeric fluid
where the stress on each particle consisting of a large number of
molecules  
depends on history of shear strain and/or shear rate.
We have investigated the simplest case of polymeric flow problems, 
and then obtained the desirable results.

Our multiscale simulation has extensibilities 
to deal with more complex phenomena on polymeric fluid
by upgrading the dumbbell model to either
a direct polymer simulation 
such as the Kremer-Grest model\cite{Krem90}
or a more effective model 
to introduce the idea of polymer reptation 
and entanglements such as
the slip-link model\cite{Shan01} 
and the primitive chain network model\cite{Masu01}.
Our application to flow problems on complex boundaries, 
e.g. contraction/expansion flow 
or free boundary flow,
can be straightforward. 
Microscopic level simulations are independent of others, 
and therefore they can be highly parallelized.

Although we referred only to polymeric fluids in the present paper,
our multiscale approach can be applied
to general flow problems, 
e.g. granular materials and powder systems,
only if the density of materials is sufficiently uniform in the
macroscopic scale in the framework of this paper.
When we apply our multiscale simulation to nonuniform density case, 
we need to delete the incompressibility condition at macroscopic level,
and
we should consider the grand canonical ensemble 
at microscopic level simulations
in order to exchange microscopic internal freedom degrees 
among macroscopic particles.

\section*{Acknowledgements}
We would like to thank 
all members in the group of {\it Multiscale Simulation for
Softmatters} for their advices and fruitful discussions.
This research was supported by 
{\it Core Research for Evolutional Science and Technology (CREST)
of
Japan Science and Technology Agency Corporation (JST)}.


\begin{thebibliography}{99}
\bibitem{Boge93} 
Boger, D. V.; Walters, K. 
Rheological phenomena in focus; 
Elsevier: Amsterdam ; London, 1993.
\bibitem{Bird89} 
Bird, R. B.; Armstrong, R. C.; Hassager, O. 
Dynamics of polymeric liquids; 
Wiley: New York ; Chichester, 1987.
\bibitem{Laso93} 
Laso, M.; \"Ottinger, H. 
JOURNAL OF NON-NEWTONIAN FLUID MECHANICS 1993, 47, 1-20.
\bibitem{Phil99} 
Phillips, T.; Williams, A. 
JOURNAL OF NON-NEWTONIAN FLUID MECHANICS 1999, 87, 215-246.
\bibitem{Ren05} 
Ren, W.; Vanden-Eijnden, E.; Maragakis, P.; E, W. 
JOURNAL OF CHEMICAL PHYSICS 2005, 123, 134109.
\bibitem{Yasu08} 
Yasuda, S.; Yamamoto, R. 
PHYSICS OF FLUIDS 2008, 20, 113101.
\bibitem{Laso97} 
Laso, M.; Picasso, M.; Ottinger, H. 
AICHE JOURNAL 1997, 43, 877-892.
\bibitem{Hali98} 
Halin, P.; Lielens, G.; Keunings, R.; Legat, V. 
JOURNAL OF NON-NEWTONIAN FLUID MECHANICS 1998, 79, 387-403.
\bibitem{Zhan04} 
Zhang, G.; Batra, R. 
COMPUTATIONAL MECHANICS 2004, 34, 137-146.
\bibitem{Mona05} 
Monaghan, J. 
REPORTS ON PROGRESS IN PHYSICS 2005, 68, 1703-1759.
\bibitem{Elle03} 
Ellero, M.; Espanol, P.; Flekkoy, E. 
PHYSICAL REVIEW E 2003, 68, 041504.
\bibitem{Doi86} 
Doi, M.; Edwards, S. F. 
The theory of polymer dynamics; Clarendon Press: Oxford, 1986.
\bibitem{Shan01} 
Shanbhag, S.; Larson, R.; Takimoto, J.; Doi, M. 
PHYSICAL REVIEW LETTERS 2001, 87, 195502.
\bibitem{Masu01} 
Masubuchi, Y.; Takimoto, J.; Koyama, K.; 
Ianniruberto, G.; Marrucci, G.; Greco, F. 
JOURNAL OF CHEMICAL PHYSICS 2001, 115, 4387-4394.
\bibitem{Kosh96} 
Koshizuka, S.; Oka, Y. 
NUCLEAR SCIENCE AND ENGINEERING 1996, 123, 421-434.
\bibitem{Otti96} 
\"Ottinger, H. C. 
Stochastic processes in polymeric fluids : 
tools and examples for developing simulation algorithms; 
Springer: Berlin ; New York, 1996.
\bibitem{Lars99} 
Larson, R. G. 
The structure and rheology of complex fluids; 
Oxford University Press: New York ; Oxford, 1999.
\bibitem{Bran98} 
Branka, A.; Heyes, D. 
PHYSICAL REVIEW E 1998, 58, 2611-2615.
\bibitem{Rapa04} 
Rapaport, D. C. 
The art of molecular dynamics simulation; 
Cambridge University Press: Cambridge, 2003.
\bibitem{Krem90} 
Kremer, K.; Grest, G. 
JOURNAL OF CHEMICAL PHYSICS 1990, 92, 5057-5086.

\end{thebibliography}
\bibliographystyle{plain}

\pagebreak

\begin{table}
\caption{Fixed parameter values in this paper.}
\label{tab_param}
\begin{center}
\begin{tabular}{cc}
\hline
$t_0$ & 1.0 [T]\\
$a$ & 1.0 [L]\\
$m$ & 1.0 [M]\\
$\eta$&$m/(a t_0)$ \\
$\Delta t$ & $1.0\times10^{-2}t_0$\\
$\rho_0$ & $m/a^3$\\
$(L_x,L_y)$&$(20a,30a)$\\
$N$ & 600\\
$h$ & $2a$\\
$l_{\rm w}$ & $3a$ \\
$\bol{f}$&$(1.0\times10^{-3},0.0)\times m/(a^2 t_0^2)$\\
\hline
\end{tabular}
\end{center}
\end{table}

\end{document}